\documentclass[%
 preprint,superscriptaddress,
showpacs,preprintnumbers,
 amsmath,amssymb,
 aps,
 prc,
]{revtex4-1}
\usepackage{graphicx}% Include figure files
\usepackage{dcolumn}% Align table columns on decimal point
\usepackage{bm}% bold math
\usepackage{subfigure}
\usepackage{multirow} %bsh hu
\usepackage{mathrsfs} %bsh hu
\begin{document}

%\begin{CJK*}{GBK}{song}
%\preprint{APS/123-QED}

\title{Gogny-force derived effective shell-model  Hamiltonian}

\author{W. G. Jiang}
\affiliation{State Key Laboratory of
Nuclear Physics and Technology, School of Physics, Peking University, Beijing 100871,
China}
\author{B. S. Hu}
\affiliation{State Key Laboratory of
Nuclear Physics and Technology, School of Physics, Peking University, Beijing 100871,
China}
\author{Z. H. Sun}
\affiliation{State Key Laboratory of
Nuclear Physics and Technology, School of Physics, Peking University, Beijing 100871,
China}
\author{F. R. Xu}\thanks{frxu@pku.edu.cn}
\affiliation{State Key Laboratory of
Nuclear Physics and Technology, School of Physics, Peking University, Beijing 100871,
China}


\date{\today}

\begin{abstract}
The density-dependent finite-range Gogny force has been used to derive the effective Hamiltonian for the shell-model calculations of nuclei. The density dependence simulates an equivalent three-body force, while the finite range gives a Gaussian distribution of the interaction in the momentum space and hence leads to an automatic smooth decoupling between low-momentum and high-momentum components of the interaction, which is important for finite-space shell-model calculations. Two-body interaction matrix elements, single-particle energies and the core energy of the shell model can be determined by the unified Gogny force. The analytical form of the Gogny force is advantageous to treat cross-shell cases, while it is difficult to determine the cross-shell matrix elements and single-particle energies using an empirical Hamiltonian by fitting experimental data with a large number of matrix elements. In this paper, we have applied the Gogny-force effective shell-model Hamiltonian to the {\it p}- and {\it sd}-shell nuclei. The results show good agreements with experimental data and other calculations using empirical Hamiltonians. The experimentally-known neutron drip line of oxygen isotopes and the ground states of typical nuclei $^{10}$B and $^{18}$N can be reproduced, in which the role of three-body force is non-negligible. The Gogny-force derived effective Hamiltonian has also been applied to the cross-shell calculations of the {\it sd}-{\it pf} shell.

%\begin{description}
%\end{description}
\end{abstract}

\pacs{21.60.Cs, 21.30.Fe, 21.10.-k}% PACS, the Physics and Astronomy
\maketitle
%\end{CJK*}
%\tableofcontents
\section{\label{sec:level1} Introduction}
Though modern {\it ab initio} methods based on realistic nuclear forces can give more fundamental understandings of strongly-interacting nuclear systems, calculations with phenomenological or empirical interactions are still popular and useful, particularly for large-scale calculations of nuclei. Among various nuclear structure calculations, the shell model is a fundamental and powerful theoretical tool which can describe the properties of both ground and excited states through the configuration mixing.

In shell-model calculations, an initial and important task is to build the effective Hamiltonian for a truncated model space. There have been several approaches to obtain the effective interactions. One common method is to start from a realistic nuclear force and use perturbation approximations to derive interaction matrix elements \cite{kuo1990folded,SUN2017227}. However, such calculation is complicated, and quantitative descriptions would require the inclusion of high-order correlations. Another one is an empirical method in which interaction two-body matrix elements (TBMEs) are  derived first from a realistic force and modified by fitting experimental data \cite{PWT,USDB,GXPF1}. However, when the model space contains two or more major shells (i.e. cross-shell cases), the fitting process becomes fairly tiring due to a large number of TBMEs. Additionally, the cross-shell single-particle energies (SPEs) are not easy to be obtained experimentally because it is difficult to distinguish single-particle excitations and collective core excitations in cross-shell cases. Usually, SPEs are input quantities in empirical shell-model calculations.

There is another way to obtain the effective shell-model Hamiltonian, that is, to use a phenomenological interaction to evaluate the TBMEs. Such calculations have been done using delta-type phenomenological interactions, e.g., the surface-delta interaction \cite{PhysRev.139.B790} and the Skyme force \cite{sagawa1985shell,gomez1993shell}. However, the delta-type force gives a constant distribution of the interaction strength in momentum space. This means that the zero-range delta-type phenomenological interactions do not provide a natural cutoff in momentum space, which has been commented in Ref. \cite{PhysRevC.81.024317}. The finite-range Gogny force \cite{GOGNY1975399,gogny_d1} gives a Gaussian distribution of the interaction in the momentum space, and hence provides a natural cutoff. The natural cutoff gives a good physics ground for the shell-model truncation in which only low-momentum components of the interaction are contained. The Gogny force has been widely used in various mean-field calculations of nuclear structure. The parameters of the Gogny force were determined by mean-field calculations fitting to the experimental data of finite nuclei and the properties of infinite nuclear matter
\cite{D1N, D1M}. The effect of the three-body force is taken into account in both the Skyrme and Gogny forces through a density-dependent term, which is essential to describe various properties of nuclei and nuclear matter. In addition, phenomenological interactions can self-consistently give the SPEs which are needed as an input of the shell-model calculation.

In the following sections, we give the detailed derivation of the shell-model effective Hamiltonian from the Gogny force. Using the effective Hamiltonian, we have made shell-model calculations for the {\it p}- and {\it sd}-shell nuclei, focusing on nuclear binding energies, excitation spectra and electric quadrupole transitions. Some cross-shell nuclei in the {\it sd}-{\it pf} space have been investigated. The calculations are compared with experimental data and other shell-model calculations with empirical or realistic-interaction Hamiltonians.

\section{\label{sec:mode} effective shell-model Hamiltonian based on the Gogny force}
In the shell-model calculation with a core, the effective Hamiltonian can be written as the sum of one- and two-body operators \cite{wildenthal1984empirical,USDB},
  \begin{eqnarray}
 H=\sum_{a}e_a  \hat{n}_a +\sum_{a\leqslant b,c\leqslant d}\sum_{JT}V_{JT}(ab;cd)\hat{T}_{JT}(ab;cd),
\label{subeq:1}
\end{eqnarray}
where the notation is standard. $e_{a}$ and $\hat{n}_a$ are the energy and particle-number operator for the single-particle orbit $a$, respectively, with the quantum numbers $(n_a,l_a,j_a)$ being the node of the radial wave function, orbital and total angular momenta, respectively.

\begin{eqnarray}
  \hat{T}_{JT}(ab;cd)=\sum_{J_zT_z}A^\dagger_{JJ_zTT_z}(ab)A_{JJ_zTT_z}(cd)
\label{subeq:2}
\end{eqnarray}
is the two-body density operator for the nucleon pair in the orbits (a, b) and (c, d) with the coupled angular momentum $J$ and isospin $T$. $A^\dagger_{JJ_zTT_z}$ or $A_{JJ_zTT_z}$ is the creation or annihilation of the nucleon pair. The establishment of the TBMEs, $V_{JT}(ab;cd)=\langle ab|V_{NN,12}|cd \rangle$, is a key step for shell-model calculations. As mentioned above, in the present paper we use the finite-range Gogny force \cite{GOGNY1975399,gogny_d1},

\begin{eqnarray}\label{gogny}
V_{NN,12}=&&\sum_{i=1}^{2}e^{-(\vec{r}_1-\vec{r}_2)^2/\mu_i^2}(W_i+B_iP^\sigma-H_iP^\tau-M_iP^\sigma P^\tau)\nonumber\\
&+&t_3\delta(\vec{r}_1-\vec{r}_2)(1+x_0 P^\sigma)\left[\rho(\frac{\vec{r}_1+\vec{r}_2}{2})\right]^{\alpha}\\
&+&iW_0\delta(\vec{r}_1-\vec{r}_2)(\vec{\sigma}_1+\vec{\sigma}_2)\cdot\vec{k}'\times\vec{k}\nonumber ,
\end{eqnarray}
where $P^\sigma=\frac{1}{2}(1+\vec\sigma_1\cdot\vec\sigma_2)$ and $P^\tau=\frac{1}{2}(1+\vec\tau_1\cdot\vec\tau_2)$ are the spin- and isospin-exchange operators, respectively, with $\vec\sigma_i$ and $\vec\tau_i$ being the spin and isospin matrix vectors, respectively. $\rho$ is the density of the nucleus at the center-of-mass (COM) position of the two interacting nucleons. The first term, with $\mu_i=0.7$ and 1.2 fm simulating two ranges of the force, gives a finite-range attraction between nucleons. The density-dependent term originates from the three-body force, generating a proper repulsive effect. The last term of Eq. (\ref{gogny}) is the spin-orbit coupling, where $\vec{k}=\frac{\overrightarrow{\nabla_1}-\overrightarrow{\nabla_2}}{2i}$ and $\vec{k'}=\frac{\overleftarrow{\nabla_1}-\overleftarrow{\nabla_2}}{2i}$ are the relative wave vectors of two nucleons, acting on the right and left sides, respectively. There have been five sets of the Gogny-force parameters which were determined by fitting experimental data and are currently used in the mean-field calculations. In the present calculations, D1S as one of the most popular Gogny interactions is used, given in Table \ref{tab:D1S_parameters}.
\begin{table}%The best place to locate the table environment is directly after its first reference in text
\caption{
\label{tab:D1S_parameters}
The D1S parameters for the Gogny force \cite{bergerD1S}.
}
\begin{ruledtabular}
\centering
\begin{tabular}{c ccccccccc p{cm}}
 &\textrm{$\mu_i$} (fm) & \textrm{$W_i$} (MeV) &\textrm{$B_i$} (MeV) & \textrm{$H_i$} (MeV) & \textrm{$M_i$} (MeV) & \textrm{$W_0$} (MeV) & \textrm{$t_3$} (MeV) & \textrm{$x_{0}$} & \textrm{$\alpha$}\\
\colrule
$i=1$ &0.7 & -1720.30 & 1300.00 & -1813.53 & 1397.60 & \multirow{2}{*}{130} & \multirow{2}{*}{1390.60} & \multirow{2}{*}{1} &\multirow{2}{*}{$1/3$} \\
$i=2$&1.2 & 103.64   & -163.48 & 162.81   & -223.93 \\
\end{tabular}
\end{ruledtabular}
\end{table}

As mentioned already in the Introduction, one of the motivations to use the Gogny force is the natural convergence when the model space increases. If we make a Fourier transformation for the finite-range term from the coordinate space to the momentum space, $e^{{-(\vec{r}_1-\vec{r}_2)}^2/\mu_i^2}$ becomes $e^{-\vec{k}^2 \mu_i^2/4}$, with $\vec{k}$ being the relative momentum of the two interacting nucleons. The Gaussian distribution makes the interaction vanish rapidly at high relative momentum. The natural decoupling between low- and high-momentum components of the interaction is important for the convergence of nuclear shell-model calculations in which the low-momentum component is dominant. The effect of three-body force is included in the form of two-body matrix elements by the density-dependent term. Earlier works using the density-dependent zero-range Skyrme force were done by Sagawa $et\ al.$ \cite{sagawa1985shell} and Gomez $et\ al.$ \cite{gomez1993shell}.

When the force is density dependent, a question arises in the diagonalization of the Hamiltonian. The shell-model calculation concerns both ground and exited states, and in principle each state corresponds to a different density. Different densities in form lead to different interactions. However it is not desired to have a different interaction for each state. In shell-model calculations based on the density-dependent Gogny force, we take the density of the ground state in the calculations of TBMEs. Such approximation has already been taken in Refs. \cite{sagawa1985shell,gomez1993shell} using the Skyrme force. In the practical calculation, the density is determined by the numerical iteration, which is similar to that in mean-field calculations. We start with harmonic-oscillator (HO) single-particle wave functions, and construct the trial configuration in which nucleons occupy the lowest HO orbits. This gives a trial density for the initial calculations of TBMEs. With the initial TBMEs, the shell-model Hamiltonian is diagonalized in the model space, and a new density of the ground state can be obtained. With the new density, we reevaluate TBMEs and diagonalize again the Hamiltonian. Such process is repeated until a converged solution is obtained. It has been tested that different starting densities always give the same result after iteration. A good trial density gives fast convergence.

The density $\rho$ in the Gogny force describes the probability of finding a nucleon at the COM position of the two interacting nucleons. In the COM coordinates, it actually relates to a local one-body density.
By the definition, the local one-body density operator in an $A$-body Hilbert space is written as \cite{PhysRevC.86.034325}
\begin{equation}
\hat{\rho} (\vec{r})=\displaystyle\sum_{k=1}^{A}\delta^{3} \left(\vec{r}- \vec{r}_{k} \right)
=\displaystyle\sum_{k=1}^{A} \frac{\delta \left(r- r_{k} \right) }{r^{2}}
\displaystyle\sum_{lm} Y_{lm}^{\ast}(\text{\^{r}}_{k}) Y_{lm}(\text{\^{r}}),
\end{equation}
where $\text{\^{r}}$ is the unit vector of $\vec{r}$,
and $Y_{lm}(\text{\^{r}})$ is the spherical harmonic function.
The density operator in the second quantization representation
within the HO basis can then be written as
\begin{eqnarray}
\hat{\rho}(\vec{r})
=&&\displaystyle\sum_{K}
\displaystyle\sum_{n' l' j' }
\displaystyle\sum_{n l j }\displaystyle\sum_{j_z}
R_{n' l' }(r)R_{n l }(r)
\frac{-Y_{K0}^{\ast}(\text{\^{r}})}{\sqrt{2K+1}}
\nonumber \\ &&\times
\left\langle l' \frac{1}{2} j' \left|| Y_{K} | \right| l  \frac{1}{2} j  \right\rangle
\left\langle j' j_z j -j_z | K 0  \right\rangle
\nonumber \\ &&\times
 (-1)^{j+j_z}
a_{n' l' j' j_z}^{\dag}a_{n l j j_z},
\end{eqnarray}
with
\begin{eqnarray}
\left\langle l' \frac{1}{2} j' \left|| Y_{K} | \right| l \frac{1}{2} j  \right\rangle
=&& \frac{1}{\sqrt{4 \pi}}  \hat{j}' \ \hat{j} \ \hat{l}' \ \hat{l}  (-1)^{j'+\frac{1}{2}}
\left\langle l' 0 l 0 | K 0  \right\rangle
\nonumber \\ &&\times
\left\{
  \begin{array}{ccc}
    j' & j &  K\\
    l & l' &  \frac{1}{2}\\
  \end{array}
\right\},
\end{eqnarray}
where $R_{n l}(r)$ is for the standard HO radial wave functions, and $K$ denotes the order of multipole expansions for the density $\rho$.
Because we are dealing with a spherically symmetric system (K=0), the one-body density operator becomes,
\begin{equation}\label{ro}
\hat{\rho} (\vec{r})=\displaystyle\sum_{n' n}\displaystyle\sum_{ljj_z}
\left[ \frac{R_{n' l}(r)R_{n l}(r)}{4 \pi} \right]
a_{n' lj j_z}^{\dag}a_{n lj j_z}.
\end{equation}
With Eq.\eqref{ro} the density distribution of the ground state (a mixed configuration) can be obtained by $\rho(\vec{R})=\langle \Phi _0|\hat{\rho}(\vec{R}) |\Phi _0 \rangle$ with ${\vec {R}}=(\vec{r}_1+\vec {r}_2)/2$ for the COM position of the two interacting nucleons. .

\subsection{\label{sec:TBME} Two-body matrix elements}
The Gogny force is written in the relative coordinate $(\vec{r}_1-\vec{r}_2)$ and center-of-mass coordinate $(\vec{r}_1+\vec{r}_2)/2$. Therefore, it is natural to derive TBMEs in the relative and center-of-mass coordinates. This can be done by using the Moshinsky transformation with the basis wave functions \cite{Moshinsky2,Moshinsky1},
\begin{eqnarray}\label{moshinsky}
|(n_{a}l_{a}m_{a})(n_{b}l_{b}m_{b})\lambda \mu  \rangle=\sum_{nlNL}M_\lambda(nlNL;n_{a}l_{a}n_{b}l_{b}) \ | (n l m)(N L M ) \lambda \mu  \rangle ,
\end{eqnarray}
where $\lambda$ denotes the total orbital angular momentum and $\mu$ is its $z$-component. $M_{\lambda}(n l N L ;n_{a}l_{a}n_{b}l_{b} )$ is the transformation coefficient called the  Moshinsky bracket \cite{Moshinsky2,Moshinsky1}. The transformation transforms the relative motion ($nlm$) and center-of-mass motion ($NLM$) of the two-particle system into two independent HO motions.

The two-body antisymmetric wave functions which we are dealing with are in the $j$-$j$ coupling scheme, therefore we need to transform the $L$-$S$ coupling into  the $j$-$j$ scheme by using the following transformation \cite{lawson1980theory},
\begin{eqnarray}\label{jj_ls}
|(n_{a}l_{a}j_{a})(n_{b}l_{b}j_{b}) J J_z  \rangle = \displaystyle\sum_{\lambda S} \displaystyle\sum_{\mu S_{z}} \gamma^{(J)}_{\lambda S}(j_{a}l_{a};j_{b}l_{b}) \langle \lambda \mu S S_{z} | J J_z\rangle   |(n_{a}l_{a}m_{a})(n_{b}l_{b}m_{b})\lambda \mu  \rangle | S S_{z} \rangle ,
\end{eqnarray}
with
\begin{eqnarray}
\gamma^{(J)}_{\lambda S}(j_{a}l_{a};j_{b}l_{b})= \sqrt{(2j_{a}+1)(2j_{b}+1)(2S+1)(2\lambda+1)}
\left\{
  \begin{array}{ccc}
    l_{a} & 1/2  &  j_{a}\\
    l_{b} & 1/2 &  j_{b}\\
    \lambda & S &  J \\
  \end{array}
\right\},
\end{eqnarray}
where $J$ and $J_z$ are the total angular momentum and its $z$-component, respectively, and $| S S_{z} \rangle $ is the two-particle spin eigenstate with $S$ being the total intrinsic spin and $S_{z}$ its $z$-component, respectively. The coefficient $\gamma^{(J)}_{\lambda S}$ is the transformation coefficient from the $L$-$S$ to $j$-$j$ schemes. The notation is standard \cite{lawson1980theory,heyde1990nuclear}, including the 9$j$ coefficients and Clebsch-Gordan coefficients.

With Eqs. \eqref{moshinsky} and \eqref{jj_ls}, a two-particle configuration in the laboratory coordinate, which couples to the quantum numbers $(JJ_{\text z}T)$, can be written as
\begin{eqnarray}\label{two-body_right}
&& \mid(n_{a}l_{a}j_{a})(n_{b}l_{b}j_{b}) J J_z T \ \rangle
 \nonumber\\
 &=&\
 \displaystyle\sum_{nlNL} \displaystyle\sum_{m M} \displaystyle\sum_{\mu S_z}\displaystyle\sum_{\lambda S} \gamma^{(J)}_{\lambda S}(j_{a}l_{a};j_{b}l_{b})
\frac{1-(-1)^{S+T+l}}{\sqrt{2(1+\delta_{n_{a}n_{b}}\delta_{l_{a}l_{b}}\delta_{j_{a}j_{b}})}}
\nonumber\\
&& \times \ M_{\lambda}(n l N L ;n_{a}l_{a}n_{b}l_{b})  \ \langle l m L M|\lambda \mu \rangle \ \langle \lambda \mu S S_{z} | J J_z\rangle \quad\nonumber\\
&& \times \ | nlm\rangle \ |N L M\rangle \ | S S_{z} \rangle \ | T \rangle ,
\end{eqnarray}
where $| T \rangle$ is the two-particle isospin eigenstate with a total isospin $T$. Then, the TBMEs in the basis given by Eq.\eqref{two-body_right} are obtained by
\begin{eqnarray}\label{two-body}
&&  \langle (n_{a}l_{a}j_{a})(n_{b}l_{b}j_{b}) J J_z T \ |\ V_{NN,12} \ | (n_{c}l_{c}j_{c})(n_{d}l_{d}j_{d}) J J_z T \ \rangle
 \nonumber\\
 &=&\
 \displaystyle\sum_{n'l'N'L'}\displaystyle\sum_{nlNL} \displaystyle\sum_{m' M'}\displaystyle\sum_{m M} \displaystyle\sum_{\mu' S'_z} \displaystyle\sum_{\mu S_z} \displaystyle\sum_{\lambda' S'} \displaystyle\sum_{\lambda S} \tilde{\gamma}^{(J)}_{\lambda' S'}(j_{a}l_{a};j_{b}l_{b}) \tilde{\gamma}^{(J)}_{\lambda S}(j_{c}l_{c};j_{d}l_{d})
\nonumber\\
&& \times \ M_{\lambda'}(n' l' N' L' ;n_{a}l_{a}n_{b}l_{b}) \ M_{\lambda}(n l N L ;n_{c}l_{c}n_{d}l_{d}) \nonumber\\
&& \times \  \langle l' m' L' M'|\lambda' \mu' \rangle \ \langle \lambda' \mu' S' S'_{z} | J J_z\rangle \langle l m L M|\lambda \mu \rangle \ \langle \lambda \mu S S_{z} | J J_z\rangle \quad\nonumber\\
&& \times \ \langle T| \ \langle S' S'_{z}| \ \langle N' L' M'| \ \langle  n'l'm'| \  \ V_{NN,12}  \ | nlm\rangle \ |N L M\rangle \ | S S_{z} \rangle \ | T \rangle ,
\end{eqnarray}
with $ \tilde{\gamma}^{(J)}_{\lambda S}(j_{a}l_{a};j_{b}l_{b})= \frac{1-(-1)^{S+T+l}}{\sqrt{2(1+\delta_{n_{a} n_{b}}\delta_{l_{a} l_{b}}\delta_{j_{a} j_{b}})}}\gamma^{(J)}_{\lambda S}(j_{a}l_{a};j_{b}l_{b})$,
where symbols with prime represent that they are for the left vector. Inputing the Gogny force into Eq. \eqref{two-body}, we can give more detailed derivation of TBMEs. Since the Gaussian terms of the Gogny force given in Eq. \eqref{gogny} only involve the relative coordinate $\vec{r}=\vec{r_1}-\vec{r_2}$, spin- and isospin-exchange operators, we simply need to calculate $\langle n' l' m' S' S'_z T' \;|\;V\;|\;n l m S S_z T \rangle$, here $V$ indicates either of the two terms in the summation of Eq. \eqref{gogny}. In the three-dimensional polar coordinates ($r$, $\theta$, $\phi$), we have
\begin{eqnarray}
&&\langle n' l' m' S' S'_z T \;|\;V\;|\;n l m S S T \rangle\nonumber\\
&=&\int\int\int r^2 \text{sin} \theta \; \langle n' l' m'  S' S'_z T \;| \ \vec{r} \ \rangle \ V  \ \langle \vec{r} \ |\;n l m S S_z T \rangle \; dr\; d\theta \; d\phi  .
\end{eqnarray}

To compute the matrix elements, the HO wave functions are needed,
\begin{eqnarray}
\langle \ \vec{r} \ |n l m\rangle &=&R_{nl} Y _{lm}(\text{\^{r}}) \nonumber \\
&=&\sqrt{\frac{2^{l-n+2}(2\nu)^{l+1.5}(2l+2n+1)!!}{\sqrt{\pi}[(2l+1)!!]^2n!}} r^le^{-\nu r^2}\nonumber\\
&&\times\sum^{n}_{x=0}(-1)^x 2^x \frac{n!(2l+1)!!}{x!(n-x)!(2l+2x+1)!!}(2\nu r^2)^x Y _{lm}(\text{\^{r}}) ,
\end{eqnarray}
where $R_{nl}$ is the radial component of the HO wave function, and $Y _{lm}(\text{\^{r}})$ is the angular component. $\text{\^{r}}$ is the unit vector of $\vec{r}$ and $r=|\vec{r}|$. $\nu=\frac{m_\mu\omega}{2\hbar}$ is the HO size parameter and $m_\mu$ is the reduced mass of the two interacting nucleons. For spherical harmonics, the following orthogonality relation is applied,
\begin{eqnarray}
\int\int \sin\theta \; Y^*_{l'm'}(\text{\^{r}}) Y_{lm}(\text{\^{r}}) \; d\theta \; d \phi=\delta_{l'l}\delta_{m'm}.
\end{eqnarray}
For spin and isospin parts, we have
\begin{eqnarray}
 \langle \ S' S'_z \ | \ P^\sigma \ | \ S S_z \ \rangle  &=&\langle \ S' S'_z \ |\ \frac{1+\vec\sigma_1\cdot\vec\sigma_2}{2}\ |\ S S_z \ \rangle \nonumber \\
 &=&  \delta_{S'S}\delta_{S'_zS_z}(S^2+S-1),
\end{eqnarray}
and
\begin{eqnarray}
 \langle \ T \ | \ P^\tau \ | \ T \ \rangle  &=&\langle \ T \ |\ \frac{1+\vec\tau_1\cdot\vec\tau_2}{2}\ |\ T \ \rangle \nonumber \\
 &=&  T^2+T-1,
\end{eqnarray}
Finally, we obtain the matrix elements of the Gaussian term,
\begin{eqnarray}
&&\langle n' l'{m}' S' S'_z T \;|\;V\;|\;n l {m} S S_z T \rangle\nonumber\\
&=&\delta_{l'l}\delta_{m'm} \delta_{S'S} \sum_{i=1}^2 \sum_{x=0}^{n'}\sum_{y=0}^{n} \ (2\nu+\frac{1}{\mu_{i}^2})^{-x-y-l'-1.5}(2x+2y+2l'+1)!! \nonumber\\
&&\times\sqrt{2^{2l'-n'+3.5}\nu^{l'+1.5}(2l'+2n'+1)!!n'!}\sqrt{2^{2l-n+3.5}\nu^{l+1.5}(2l+2n+1)!!n!}\nonumber\\
&&\times(-1)^{x+y}  2^{-l'-2+x+y} \frac{1}{(n'-x)!x!}\frac{1}{(n-y)!y!}\frac{1}{(2l'+2x+1)!!}\frac{1}{(2l+2y+1)!!} \\
&&\times[W_i+B_i(S^2+S-1)-H_i(T^2+T-1)-M_i(S^2+S-1)(T^2+T-1)]. \nonumber
\end{eqnarray}

The above derivations have been cross checked with calculations using partial-wave decompositions up to L=4 \cite{parialwave}. The TBMEs calculations of the density-dependent and spin-orbit coupling terms are similar to those in Refs. \cite{sagawa1985shell,gomez1993shell} using the Skyrme force. However, Refs. \cite{sagawa1985shell,gomez1993shell} did not perform the shell-model iteration for a self-consistent density, while a Hartree-Fock density \cite{sagawa1985shell} or an approximate HO density \cite{gomez1993shell} was used.

To test the feasibility of using the Gogny interaction to describe the shell-model effective interaction, we have compared the Gogny-D1S TBMEs with other frequently-used effective interactions.
\begin{figure}
\centering
\includegraphics[width=0.9\textwidth]{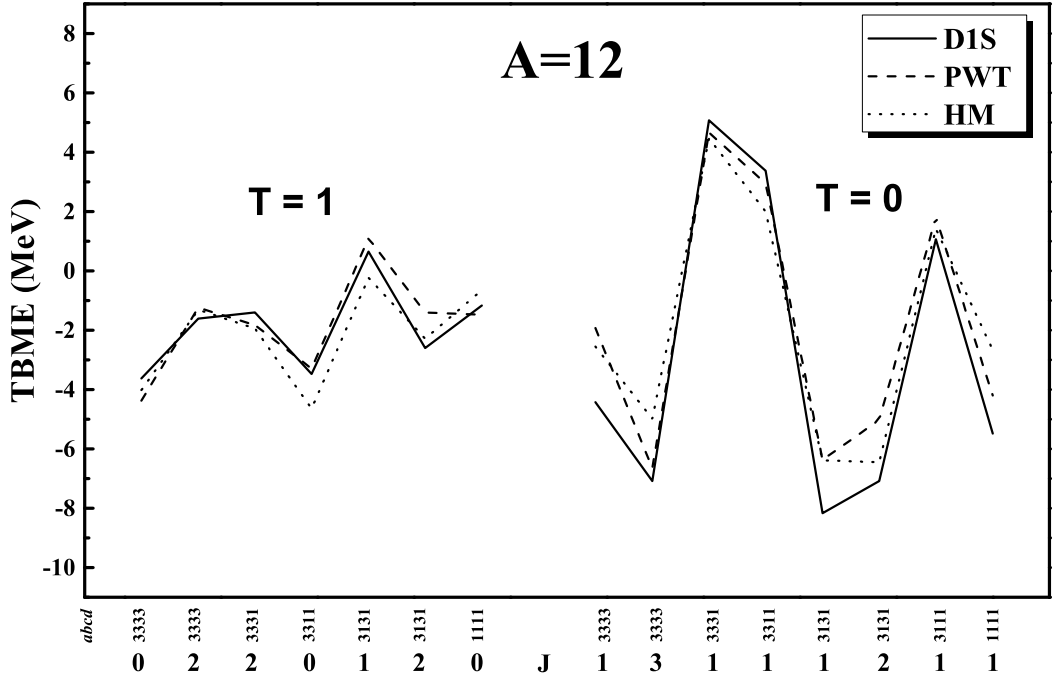}
\caption{\label{fig:p_TBME} Comparison of the Gogny D1S, empirical PWT and realistic-force HM two-body matrix elements in $j$-$j$ scheme (A=12). The $abcd$ orbits are labeled by 1\,=\,$p_{1/2}$, 3\,=\,$p_{3/2}$.}
\end{figure}
In the $p$ shell, the TBMEs of the Gogny D1S \cite{bergerD1S} are compared with the empirical PWT interaction by Warburton \cite{PWT} and the realistic-force HM matrix elements by Hauge and Maripuu \cite{PhysRevC.8.1609}. Fig. \ref{fig:p_TBME} displays the 15 TBMEs of the D1S, PWT and HM interactions in $^{12}$C for two isospin channels, showing good agreements in the three interactions.

\begin{figure}
\centering
\includegraphics[width=0.9\textwidth]{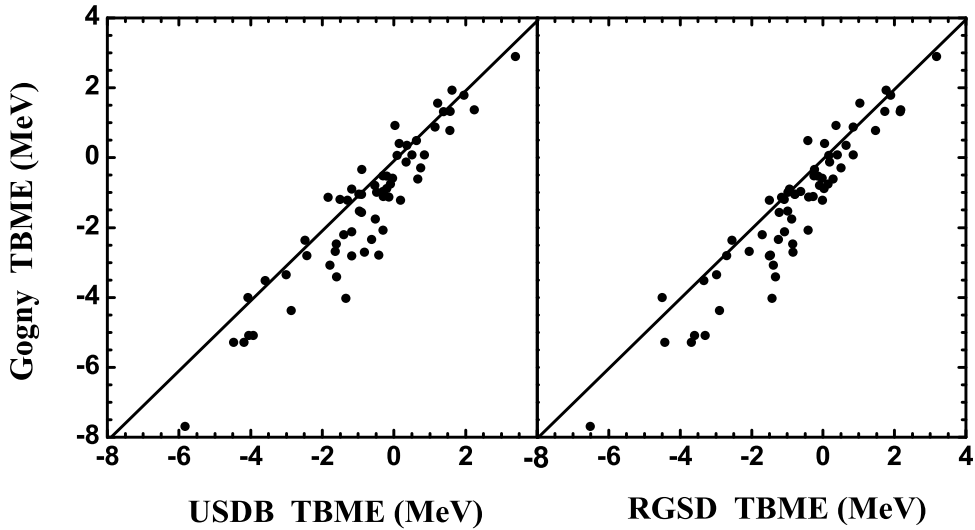}
\caption{\label{fig:sd_TBME}Comparison of D1S, USDB and RGSD two-body matrix elements (A=18).}
\end{figure}

In the $sd$ shell, the 63 matrix elements of the Gogny D1S interaction are compared with the empirical USDB interaction by Brown \cite{USDB,wildenthal1984empirical} and the realistic-force RGSD interaction by Hjorth-Jensen \cite{hjorth1995realistic}, shown in Fig. \ref{fig:sd_TBME}. The USDB interaction is obtained by using a linear-combination method iteratively (starting from RGSD) to reach a best fit towards experimental data \cite{USDB}. RGSD is derived by applying G-matrix and folded diagrams to the Bonn-A $NN$ potential. In Fig. \ref{fig:sd_TBME}, we see the Gogny TBMEs are similar to both USDB and RGSD with the rms deviations of 333 keV and 322 keV, respectively.

\subsection{\label{sec:SPE}Single-particle energies}
In addition to TBMEs, the SPEs in the model space are also important quantities for the shell-model calculation with a core. In the present paper, we calculate the SPEs by \cite{brown2005lecture}
\begin{eqnarray}\label{spe}
e_{j}=t_{j}+\frac{1}{2(2j+1)}\sum_{j^\text{c}}\sum_{JT}(2J+1)(2T+1)\langle \ jj^\text{c}JT\ | \ V \ |\ jj^\text{c}JT \ \rangle ,
\end{eqnarray}
where $j$ stands for a valence-particle orbit (with the standard quantum numbers $n,l,j$) and $j^\text{c}$ for an orbit in the core. Here, SPE is considered to be the sum of the valence-particle kinetic energy and its interaction with all the nucleons in the core.

Actually, for most effective interactions in a single shell, the SPEs are either determined by fitting data along with TBMEs or straightly taken to be experimental values. However, in cross-shell or heavy-mass case, it is difficult to determine the values of SPEs because single-particle excitations are often submerged by collective core excitation and it is hard to gain sufficient experimental data. With Eq. \eqref{spe}, the present method can overcome the above problems and provide an alternative way to determine the SPEs for any given model space. Another advantage of employing Eq. \eqref{spe} is that the same Gogny force is used for the calculations of single-particle energies. Such unified treatment provides a self-consist way to obtain binding energies.

\subsection{\label{sec:GSE} Ground-state energies}
The ground-state energy of a nucleus can be written as
\begin{eqnarray}
E_{\text{g.s.}}=E_{\nu}+E_{\text{coul}}-t_{\text {COM}}+E_{\text{c}},
\end{eqnarray}
here $E_{\nu}$ stands for the valence-particle energy given by the shell-model diagonalization with the TBMEs and SPEs obtained above. $E_{\text{coul}}$ is the coulomb energy and $t_{\text {COM}}=\frac{3}{4} \hbar \omega$ is the center-of-mass kinetic energy. $E_{\text{c}}$ represents the energy of the core.

In shell-model calculations with empirical interactions, the core energy usually takes the experimental energy of the core nucleus and does not change with the mass number $A$. In the present calculation, the core energy is calculated by
\begin{eqnarray}
E_{\text{c}}=t_{\text{c}}+V_{\text{c}},
\end{eqnarray}
with the core kinetic energy by
\begin{eqnarray}
t_{\text{c}}=\sum_{j^\text{c}_a}(2T+1)(2J+1)\langle j^\text{c}_a |\hat{t}|j^\text{c}_a \rangle,
\end{eqnarray}
and the core potential by
\begin{eqnarray}
V_{\text{c}}=\sum_{j^\text{c}_a\leq j^\text{c}_b}\sum_{JT}(2T+1)(2J+1)\langle j^\text{c}_a j^\text{c}_b J T|V|j^\text{c}_a j^\text{c}_b J T \rangle .
\end{eqnarray}
$j^\text{c}_a$, $j^\text{c}_b$ stands for orbits in the core and $\hat{t}$ is the single-particle kinetic energy operator. Note that the interaction $V$ here is taken to be the same Gogny force as in TBMEs. Due to the density dependence of the Gogny force, the core energy is smoothly $A$-dependent, which has been discussed in Ref. \cite{gomez1993shell} where the density-dependent Skyrme force was used. In fact, the $A$-dependent core energy is crucial to reproduce experimental binding energies for a whole chain of isotopes.
\begin{figure}
\centering
\includegraphics[width=0.9\textwidth]{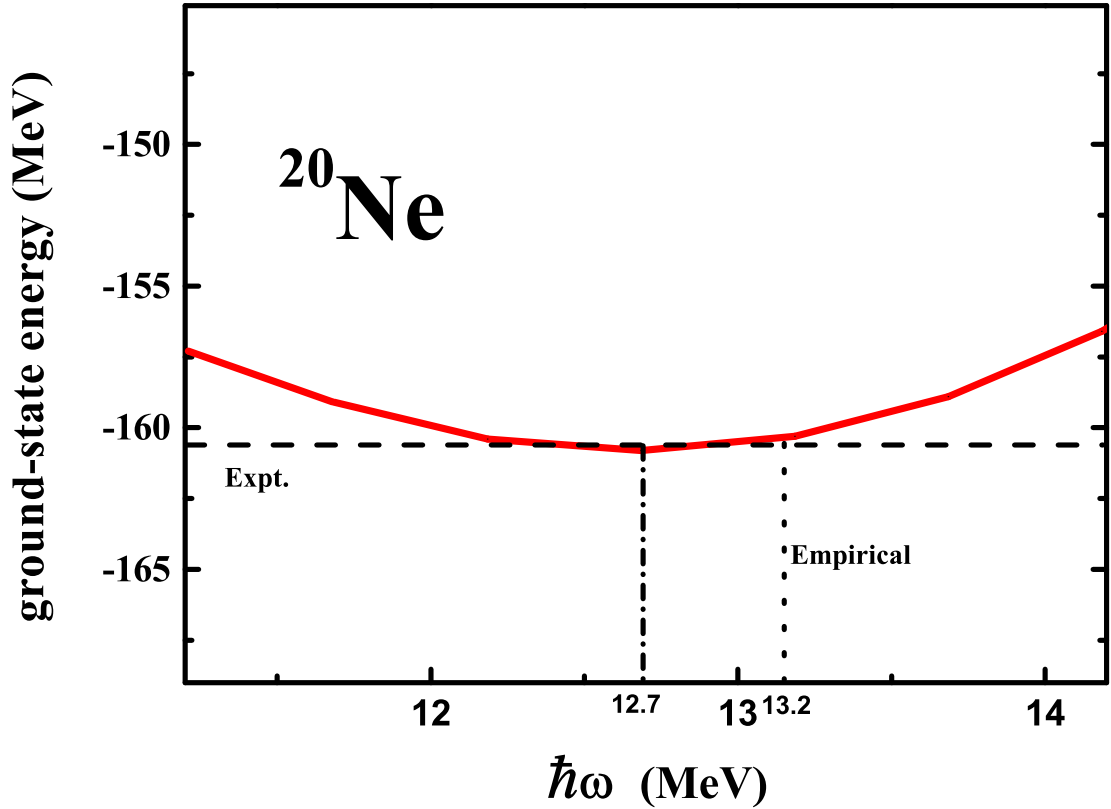}
\caption{\label{fig:ne20_hw} $^{20}$Ne ground-state energy as a function of $\hbar\omega$, calculated by the shell model with the Gogny force.}
\end{figure}

The TBMEs, SPEs and core energy are calculated by the unified Gogny interaction in a HO basis at certain $\hbar \omega$. Therefore, the calculation would be $\hbar\omega$ dependent, which has been commented in Ref. \cite{gomez1993shell}. In no core shell model \cite{PhysRevC.62.054311} or other shell models (e.g., shell model based on the Skyrme force \cite{gomez1993shell}), the $\hbar\omega$ value which minimizes the binding energy is used. In the shell-model calculation with an empirical interaction, usually the empirical $\hbar\omega=45 A^{-1/3}-25A^{-2/3}$ \cite{brown2005lecture} is adopted. We find that in the present calculations the $\hbar\omega$ parameter determined by minimizing the binding energy is close to the value given by the empirical formula. The two choices of the $\hbar\omega$ values give similar binding energies. Fig. \ref{fig:ne20_hw} shows an example of the calculation for $^{20}$Ne. The empirical $\hbar\omega$ value gives a good description of nuclear radii. In the present paper, we adopt the empirical $\hbar\omega$.

\subsection{\label{sec:CenterOfMass}Center-of-mass correction}

When the model space involves two or more HO major shells, the COM correction must be considered. The COM motion can produce a non-physics spurious excitation with a $\hbar \omega$  excitation energy or higher. In the HO basis, the spurious COM excitation can be removed using the Lawson method \cite{gloeckner1974spurious} by adding a multiplied COM Hamiltonian,   $\beta H_{\text{COM}}=\beta (\frac{(\sum_{i=1}^{A}p_i)^2}{2Am}+\frac{1}{2}\frac{m\omega^2}{A}(\sum_{i=1}^{A}r_i)^2-\frac{3}{2}\hbar\omega)  $, into the shell-model Hamiltonian, given $H'=H_{\mathrm{SM}}+\beta H_{\mathrm{COM}}$, here $\beta$ is the multiplying constant. The Lawson method actually pushes up the COM excitation energy. If the constant $\beta$ takes a large enough value, the spurious COM excitation can be separated from the intrinsic low-lying states of interest.

\section{\label{sec:Results} Ground-state energies and excitation spectra}

\begin{figure}
\centering
\includegraphics[width=0.9\textwidth]{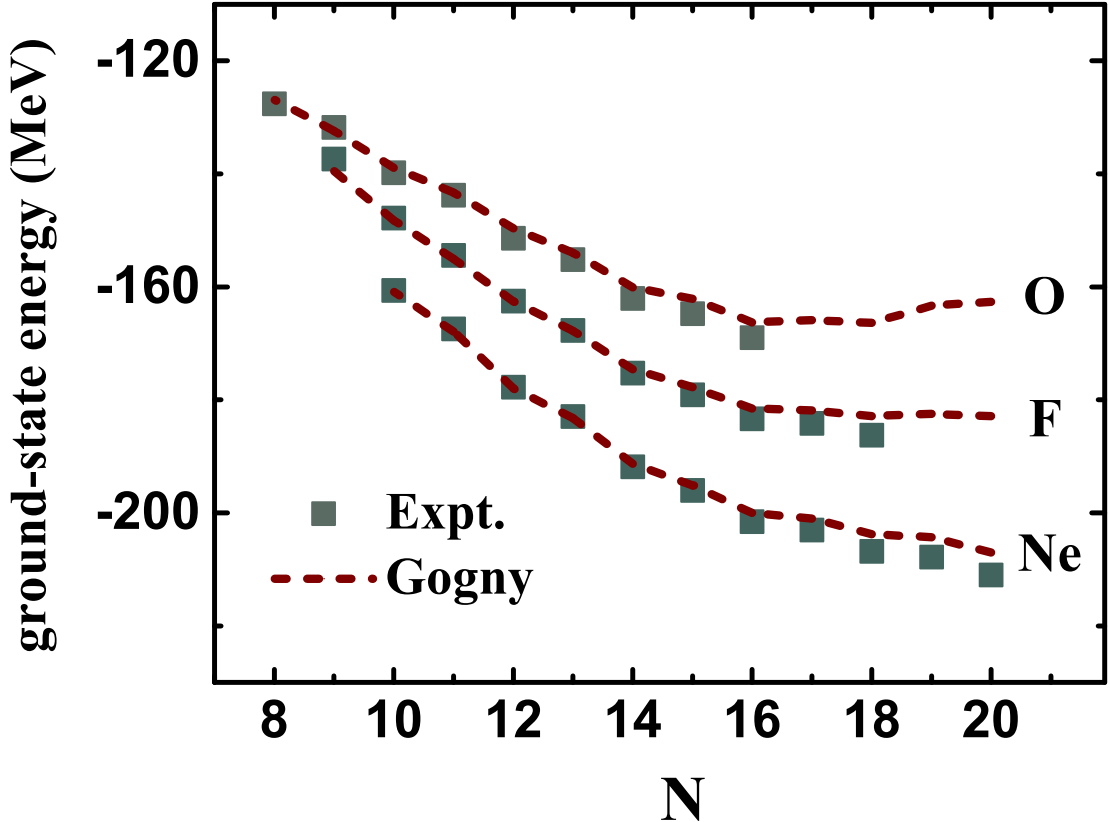}
\caption{\label{fig:OFNE_BE}(Color online) Calculated ground-state energies (dot lines) for oxygen, fluorine and neon isotopes. Experimental data (squares) are taken from Ref. \cite{nndc}. The present shell-model calculations based on the Gogny force is indicated by ``Gogny".}
\end{figure}
\begin{figure}
\centering
\includegraphics[width=0.9\textwidth]{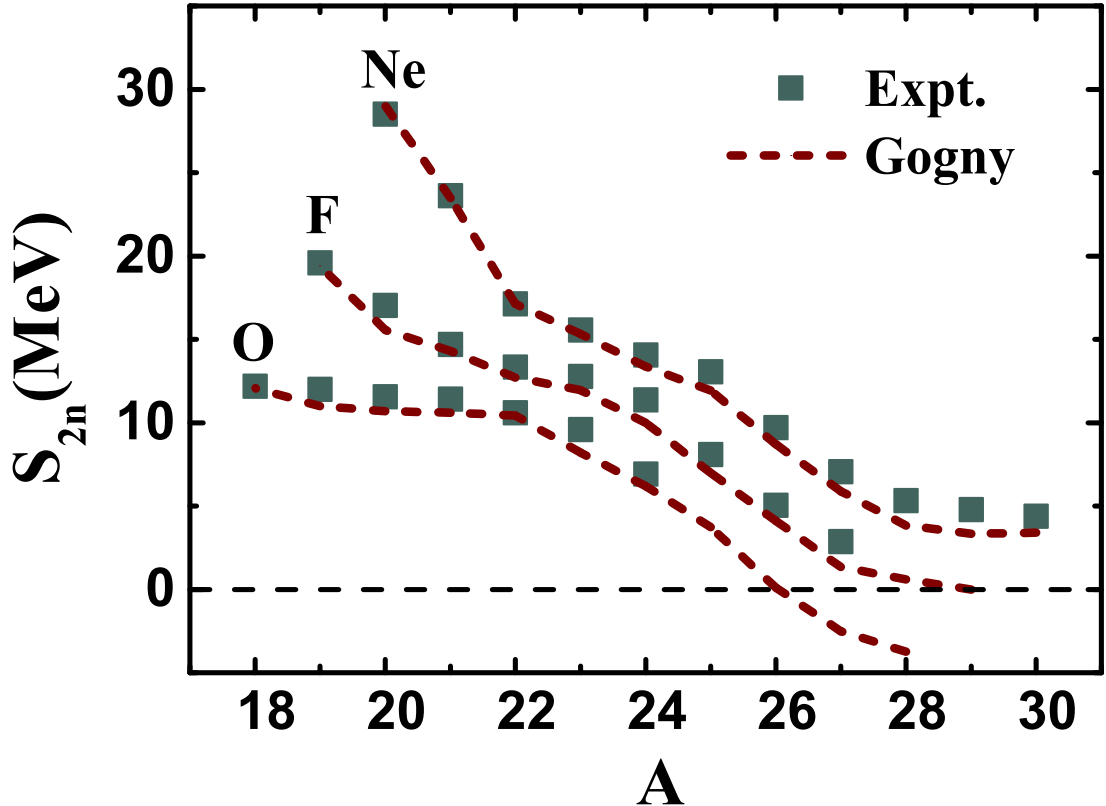}
\caption{\label{fig:OFNE_s2n}(Color online) Calculated two-neutron separation energies (dot lines) for oxygen, fluorine and neon isotopes. Experimental data (squares) are taken from Ref. \cite{nndc}.}
\end{figure}

%As discussed in Section \ref{sec:mode}, the binding energy is the key to determine the HO parameter and the start of our calculation. Therefore it is instructive to see whether Gogny force can provide reasonable ground-state energies.

Figs. \ref{fig:OFNE_BE} and \ref{fig:OFNE_s2n} present the ground-state energies and two-neutron separation energies for the O, F and Ne isotopic chains of the $sd$ shell, respectively. It shows that the Gogny-force shell model can well reproduce the experimental ground-state energies. The experimentally-known neutron drip-line position of the oxygen chain is reproduced, which is at $^{24}$O. In experiment, $^{25}$O is the first unbound nucleus behind the neutron drip-line nucleus, $^{24}$O. In our calculation, the energy of $^{25}$O is about 0.5 MeV unbound with respect to $^{24}$O. For fluorine isotopes, we reproduce the unbound nature of $^{28}$F with an unbound energy of 0.4 MeV above the threshold of neutron emission. $^{26}$O and $^{29}$F are bound against one-neutron emission. While in Fig. \ref{fig:OFNE_s2n} the negative values of two-neutron separation energies indicate that both of them are unbound with two-neutron emission. As for neon isotopes, $^{30}$Ne is still well bound and the drip line of the neon chain should be behind the $sd$ shell. There have already been many shell-model calculations for the {\it sd}-shell mass region, e.g., with a monopole-based interaction \cite{PhysRevC.85.064324}. In this paper, we focus on how well the Gogny force is applied to the shell-model calculations.

In Refs. \cite{PhysRevLett.105.032501,PhysRevLett.110.242501}, it was pointed out that empirical two-body interactions (e.g., the monopole-based interaction \cite{PhysRevC.85.064324}) or $ab \ initio$ interaction with three-body force may describe the drip line of oxygen isotopes, whereas $ab \ initio$ interactions without three-body force fails. In empirical interactions, we may assume that the effect of the three-body force is partially included in the matrix elements that are determined by fitting data. In the present calculation, the density-dependent term of the Gogny force provides a repulsion, while other terms mainly contribute to the attraction. It is the density dependence (an equivalent three-body force) that prevents the ground-state energies of isotopes from endless dropping down with increasing neutron number and gives a reasonable description of the drip line. It has been known that the empirical WBP \cite{PWT} and WBT \cite{PWT} interactions cannot correctly give the spins of the $^{10}$B and $^{18}$N ground states. In Ref. \cite{B10}, it was proved that the {\it ab initio} calculation with three-body force can well describe the ground states of $^{10}$B and $^{18}$N.
\begin{figure}
\centering
\includegraphics[width=0.9\textwidth]{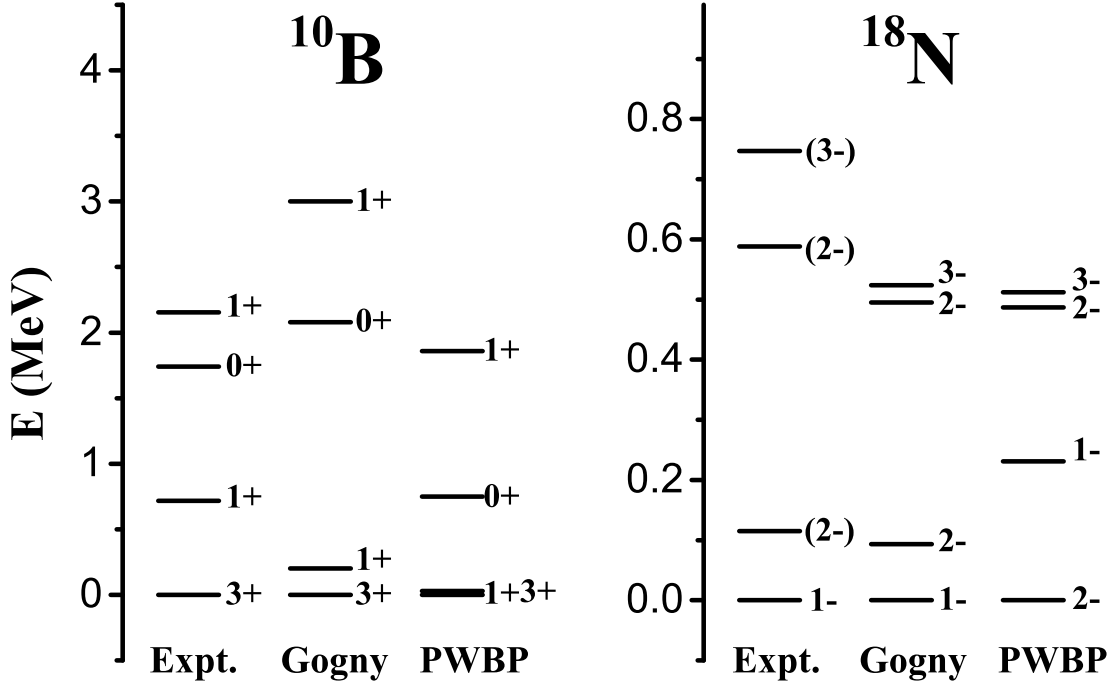}
\caption{\label{fig:b10n18}(Color online) Excitation spectra for $^{10}$B and $^{18}$N, calculated by Gogny (D1S) and the empirical interactions (PWBP, WBP), compared with experimental data \cite{nndc}.}.
\end{figure}
Fig. \ref{fig:b10n18} shows the calculated spectra of $^{10}$B and $^{18}$N by the Gogny force, compared with experimental data as well as the common empirical interactions. We see that with the inclusion of the three-body effect through the density dependence, the Gogny-force calculations give the correct ground-state properties of the two nuclei.
%good description of the low lying spectra and give the correct ground states for these nuclei.

\begin{figure}
\centering
\includegraphics[width=0.9\textwidth]{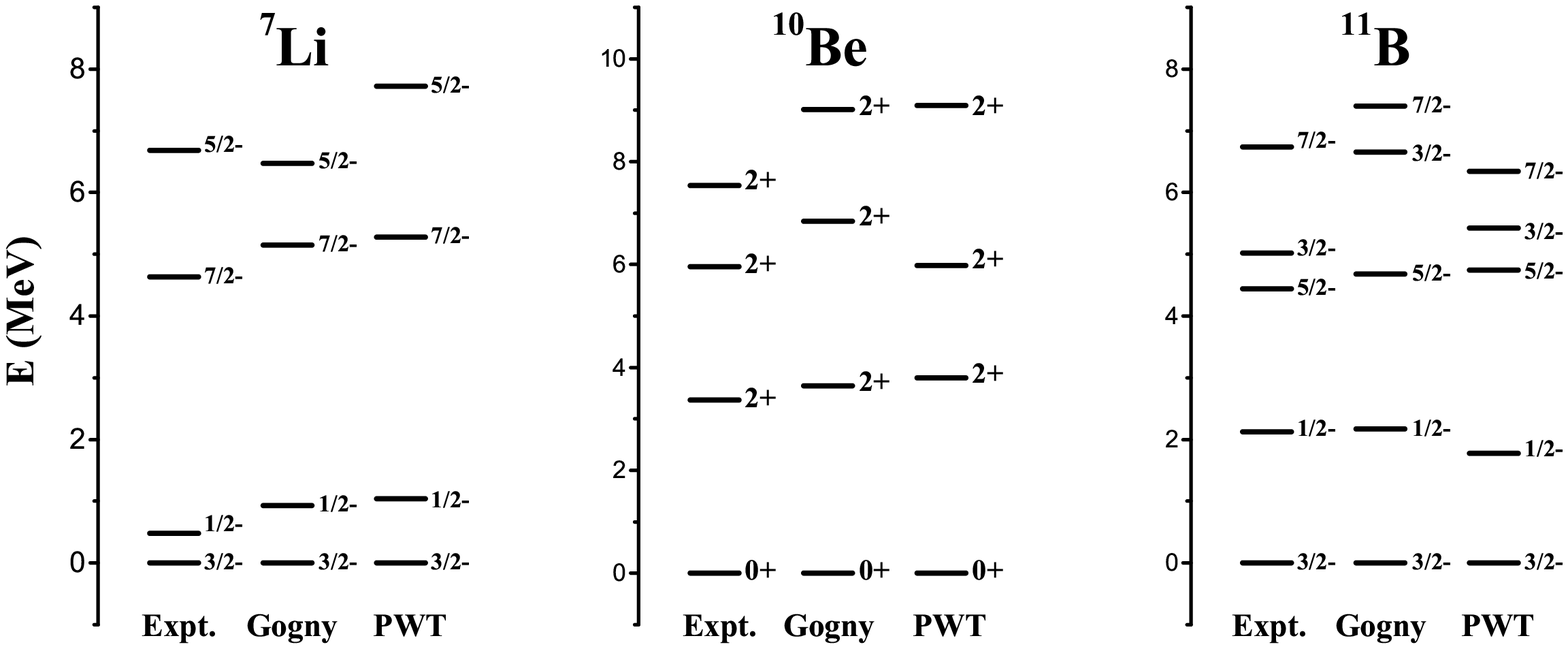}
\caption{\label{fig:li7be10b11}(Color online) Energy levels for $^{7}$Li, $^{10}$Be, $^{11}$B, calculated with the Gogny (D1S) interaction, compared with experimental data \cite{nndc} and the shell-model calculations based on the empirical PWT interaction.}.
\end{figure}
\begin{figure}
\centering
\includegraphics[width=0.9\textwidth]{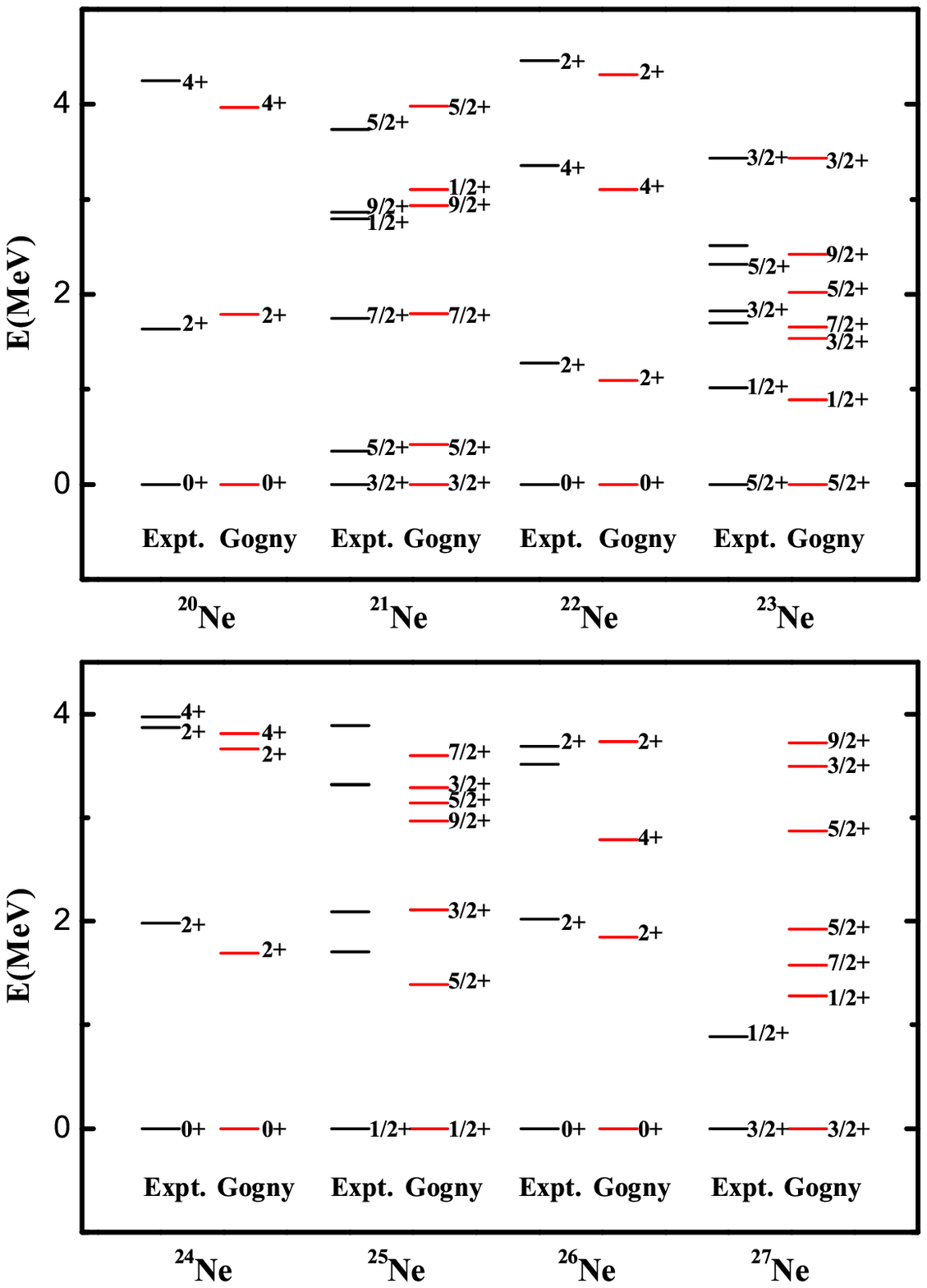}
\caption{\label{fig:ne20-27}(Color online) Energy levels for $^{20-27}$Ne, obtained in the Gogny (D1S) interaction, compared with experimental data \cite{nndc}.}.
\end{figure}

In the present paper, we have performed the Gogny-force shell-model calculations for the {\it p} and {\it sd} shells. Fig. \ref{fig:li7be10b11} presents the energy levels of $^7$Li, $^{10}$Be and $^{11}$B in the $p$ shell. The shell-model calculations using the Gogny force are in good agreements with data and those using the empirical PWT \cite{PWT} interaction. Fig. \ref{fig:ne20-27} shows the spectra of $^{20-27}$Ne in the $sd$ shell. The agreement between calculations and data is fairly good.
%For even-$A$ neon isotopes, all positive-parity levels are well matched between theory and experiment. For odd-$A$ isotopes, most of the levels are in right sequence except for $^{21}$Ne. The first $1/2^{+}$ and $9/2^{+}$ states of this nucleus are inverted compared to experiment.
%In our calculation, the gaps between the three valence orbits in $sd$ shell are slightly larger than experimental ones. The spectrum of odd-$A$ nuclei will strongly depend on SPE values since the configurations of these energy levels are dominated by the single particle excitations. It is worth mentioning that most of the effective interactions in $sd$ shell also give the same inversion for this nucleus including USDB.

\begin{figure}
\centering
\includegraphics[width=0.8\textwidth]{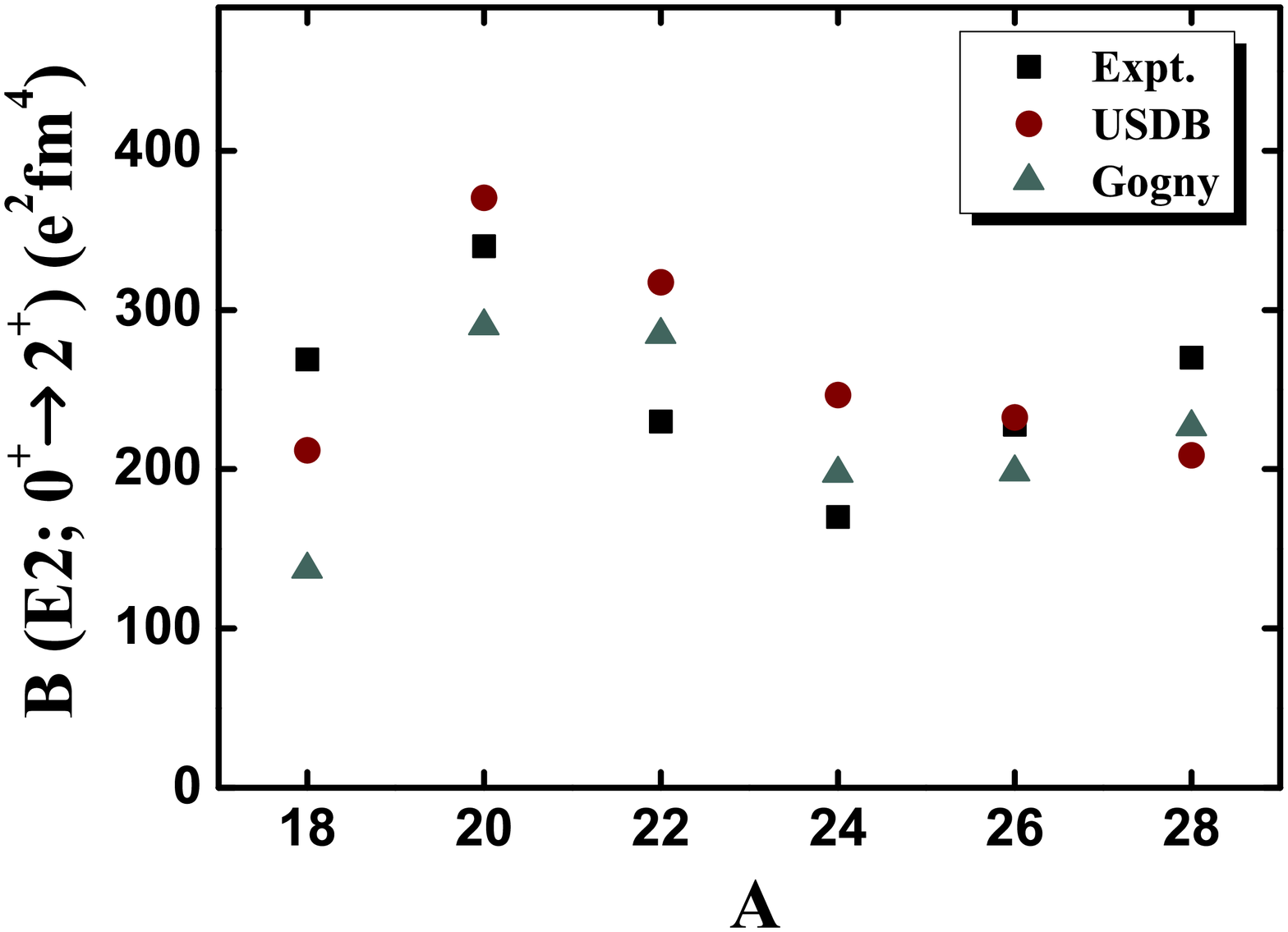}
\caption{\label{fig:ne18-28_BE2} (Color online) B(E2) values calculated for Ne isotopes with the Gogny and USDB interactions, compared with experimental data \cite{BE2}.}
\end{figure}

\begin{figure}
\centering
\includegraphics[width=0.8\textwidth]{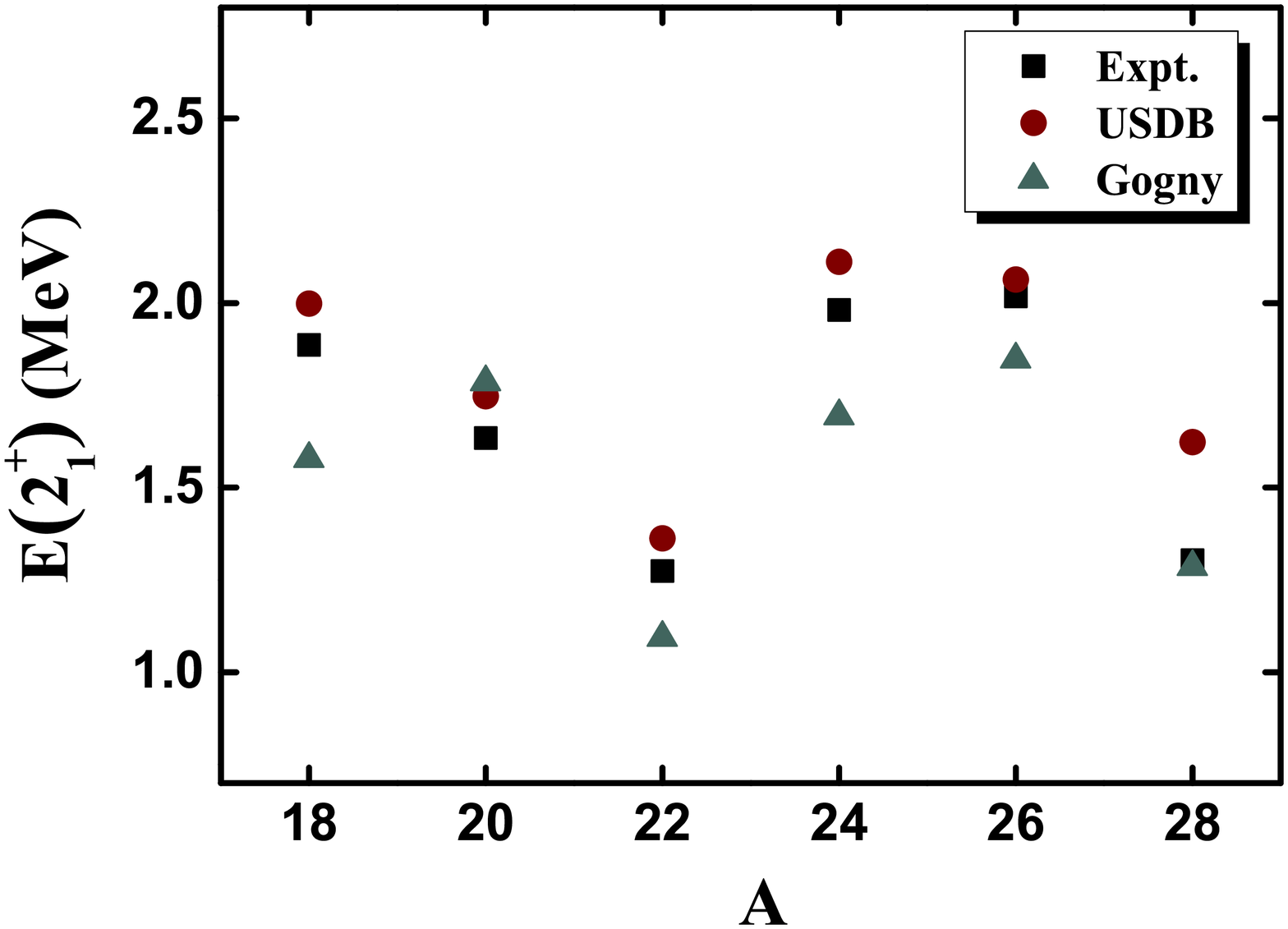}
\caption{\label{fig:ne18-28_E2} (Color online) The energies of the first $2^+$ excited states in Ne isotopes, calculated with Gogny and USDB interactions, compared with experimental data \cite{BE2}.}
\end{figure}

The electromagnetic transition between levels is another important observable in nuclear spectroscopy. In the shell model with the Gogny force, the electric quadrupole transitional probability $B(E2)$ between the ground state and the first $2^+$ in even-even nuclei has been calculated. In shell-model calculations, the model-space effective charges for the proton and neutron are usually used, which considers effects from the core polarization and excluded space. We use standard effective charges $e_{p}$=1.5 and $e_{n}$=0.5 for the $sd$-shell nuclei. Fig. \ref{fig:ne18-28_BE2} shows the calculated $B(E2; 0^{+}\rightarrow2^{+})$ for Ne isotopes using the Gogny interaction. For comparison, experimental data as well as calculated $B(E2)$ values by the USDB interaction are included. It is clear that the tendency of $B(E2)$ along the Ne isotopic chain is in agreement with data for both the Gongy and USDB calculations. To see the systematics, we have also plotted the energies of the first $2^+$ excited states of the Ne isotopes in Fig. \ref{fig:ne18-28_E2}. The large $2^+_1$ excitation energies at $^{24}$Ne and $^{26}$Ne indicate neutron sub-shell closures at $N=14$ and 16 in the neutron-rich neon isotopes. A larger $B(E2)$ value implies more collective.

One of the advantages using a phenomenological interaction is that the shell model can easily go to cross-shell calculations, while it is difficult to obtain TBMEs and SPEs in the empirical method by fitting data. Using the unified Gogny force, we can simultaneously calculate both TBMEs and SPEs for a cross-shell space which contains two or more major shells. Nuclei locating the $N\sim20$ ``island of inversion'' are in one of the typical cross-shell regions \cite{PhysRevC.41.1147}. The nuclei around this region own many anomalous properties, such as remarkably low $E(2^+_1)$ and large $B(E2)$ \cite{PhysRevC.12.644,huber1978spins,detraz1979beta,motobayashi1995large}, which indicates probably the quenching of the $N=20$ shell gap and the intrusion of the {\it pf} neutron orbits. In this situation, the single-shell shell-model calculation is no longer applicable \cite{campi1975shape}, and cross-shell calculations should be needed. For example, the large-scale shell model by Caurier \emph{et al.} \cite{caurier2001shell} and Monte Carlo shell model by Utsuno \emph{et al.} \cite{utsuno2002monte} have been performed. To explain the recent experiment on the $\beta$ decays of the $N\sim 20$ "island of inversion" nuclei \cite{HAN_1}, we performed some calculations using the Gogny force for the nuclei in the mass region. The results are in good agreements with the experimental data \cite{HAN_1} and the calculations using the SDPF-M interaction \cite{SDPF-M_1,SDPF-M_2}. In the present paper, we calculate excitation spectra for the neutron-rich neon isotopes in and around the $N\sim 20$ "island of inversion". The main purpose of the paper is to give the detailed formulation and test the general feasibility of the shell model with the Gogny force.

\begin{figure}
\centering
\includegraphics[width=1\textwidth]{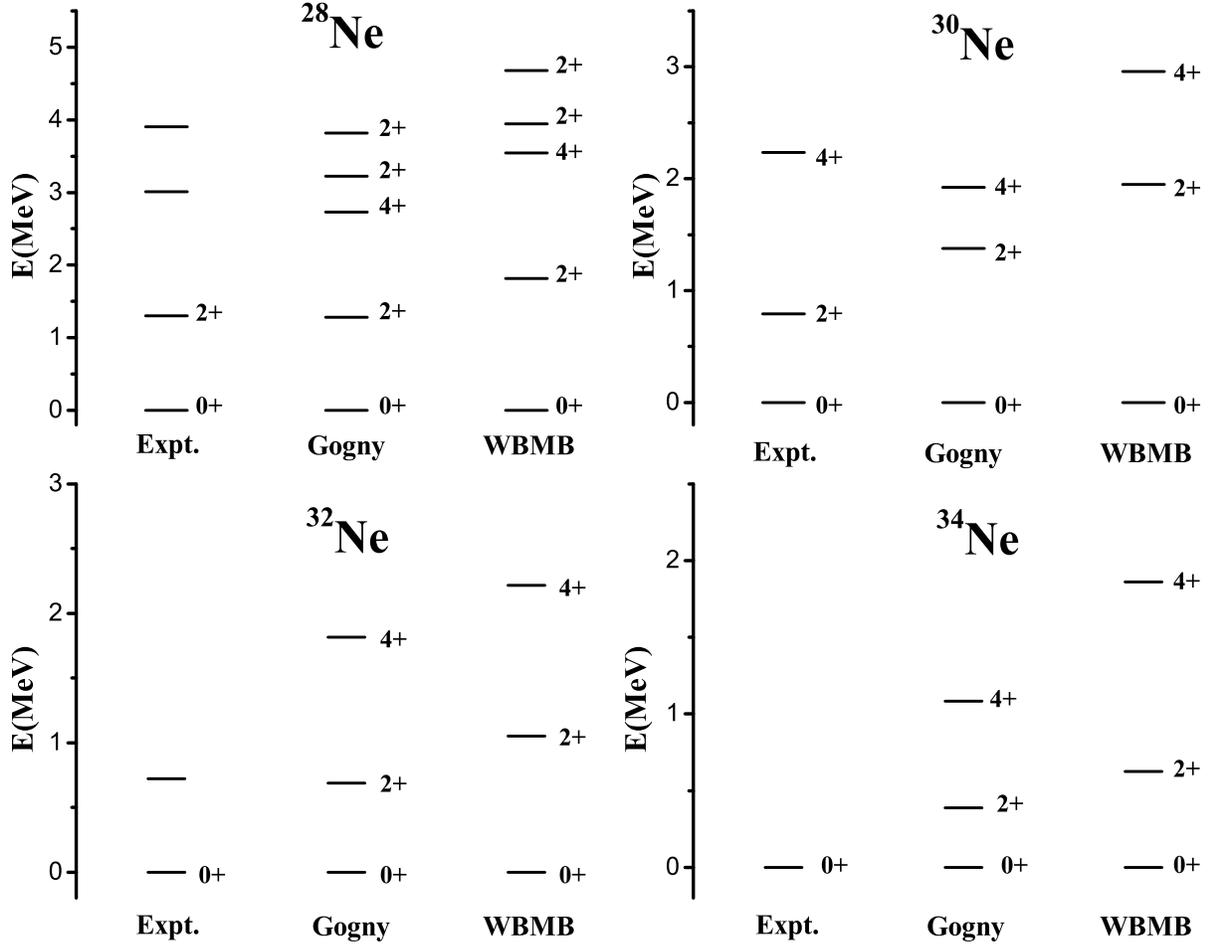}
\caption{\label{fig:Ne28-34}  Cross-shell calculations for the excitation spectra of even-even $^{28-34}$Ne isotopes, by the phenomenological Gogny and empirical WBMB interactions, compared with experimental data \cite{nndc}.}
\end{figure}

The model space is protons in the {\it sd} shell and neutrons in the {\it sd}-{\it pf} ($f_{7/2}p_{3/2}$) shell which is frequently used \cite{PhysRevLett.80.2081,PhysRevLett.77.3967}. In practical calculations, we keep the neutron $d_{5/2}$ shell occupied fully. This is to reduce the model dimension and reasonable, because the $d_{5/2}$ orbits are bound deeply. The effect from the $d_{5/2}$ excitation should be less important for lowly-excited states. Fig. \ref{fig:Ne28-34} shows the low excitation spectra of the nuclei, $^{28}$Ne, $^{30}$Ne, $^{32}$Ne, $^{34}$Ne, compared with data and the calculations with the empirical WBMB interaction \cite{PhysRevC.40.2823}. The WBMB interaction includes the $sd$-shell TBMEs by Wildenthal \cite{wildenthal1984empirical}, the $pf$-shell TBMEs by McGory \cite{mcgrory1973shell}, and a modified Millener-Kurath \cite{millener1975particle} interaction for the cross-shell matrix elements. The Gogny calculation gives lower energies for the excited states than WBMB, which indicates stronger mixtures in cross-shell configurations. For higher-energy excited states, one needs to increase the configuration space with the neutron $d_{5/2}$ shell unfrozen. The spurious center-of-mass excitation has been treated using the Lawson method \cite{gloeckner1974spurious} that has been described above. Our $\beta$-decay calculations for the $sd$-$pf$ region can be found in the recent experimental paper \cite{HAN_1}.

\section{Summary}
We have derived an effective shell-model Hamiltonian based on the finite-range density-dependent Gogny force. The detailed formulation is given in the paper. The finite range gives a natural cutoff of the interaction between low momentum and high momentum. The density dependence which originates from the three-body force plays a crucial role in predicting nuclear drip lines and describing the property of the ground states of $^{10}$B and $^{18}$N. The density distribution that appears in the interaction is determined self-consistently by employing the iteration with the shell-model diagonalizing. In a given model space, single-particle energies and effective two-body matrix elements are calculated using the unified Gogny interaction.

The Gogny force allows us to calculate the energy of the core. The $A$-dependent core energy is crucial in describing the binding energies of isotopes of the whole chain and predicting the drip lines. We have applied the Gogny-force shell model to the {\it p}-shell and {\it sd}-shell nuclei. The model can well describe the excitation spectra and the electric quadrupole transitional probabilities of the nuclei. The binding energies of oxygen, fluorine and neon isotopes are well reproduced.

The phenomenological Gogny force provides an easy way to calculate cross-shell interaction matrix elements and single-particle energies, which is difficult in the empirical method by fitting data. As example, we have investigated some neutron-rich neon isotopes in the {\it sd}-{\it pf} shell. Satisfactory results are obtained. Further calculations will be done in future papers.

\begin{acknowledgments}
Valuable discussions with Calvin Johnson are gratefully acknowledged. This work has been supported by the National Key R$\&$D Program of China under Grant No. 2018YFA0404401; the National Natural Science Foundation of China under Grants No. 11835001, No. 11320101004 and No. 11575007; the China Postdoctoral Science Foundation under Grant No. 2018M630018; and the CUSTIPEN (China-U.S. Theory Institute for Physics with Exotic Nuclei) funded by the U.S. Department of Energy, Office of Science under Grant No. DE-SC0009971. We acknowledge the High-performance Computing Platform of Peking University for providing computational resources.
\end{acknowledgments}

\bibliography{references}
\end{document}